\def\Msun{M_{\odot}}
\def\Lsun{L_{B,\odot}}

\def\kms{{\rm\, km\,s^{-1}}}
\def\ergs{{\rm\, erg\,s^{-1}}}

\def\Tvir{T_{\rm vir}}

\def\Mgal{M_{\rm gal}}

\def\DMinf{{\dot M}_{\rm inf}}
\def\tinf{\tau_{\rm inf}}

\def\Mgas{M_{\rm gas}}
\def\DMgas{{\dot M}_{\rm gas}}

\def\DMrec{{\dot M}_{\rm rec,*}}

\def\Mstar{M_*}
\def\DMstar{{\dot M}_*}
\def\Wstar{W_*}
\def\alstar{\alpha_*}
\def\tdyn{\tau_{\rm dyn}}
\def\tcool{\tau_{\rm cool}}

\def\DMesc{{\dot M}_{\rm esc}}
\def\tesc{\tau_{\rm esc}}

\def\Mbh{M_{\rm BH}}
\def\DMbh{{\dot M}_{\rm BH}}
\def\DMbhac{{\dot M}_{\rm BH,acc}}

\def\bhstar{\beta_{\rm BH,*}}
\def\DMedd{{\dot M}_{\rm Edd}}
\def\DMbon{{\dot M}_B}

\def\fed{f_{\rm Edd}}
\def\Ledd{L_{\rm Edd}}
\def\Lbh{L_{\rm BH}}
\def\Lb{L_B}
\def\Lx{L_X}

\def\etal{{et al.}}
\def\mnr{\textit{MNRAS}}
\def\apj{\textit{ApJ}}
\def\aj{\textit{AJ}}
\def\aea{\textit{A\&A}}
\def\araa{\textit{ARAA}}

\catcode`\@=11
\def\gsim{\ifmmode{\mathrel{\mathpalette\@versim>}}
    \else{$\mathrel{\mathpalette\@versim>}$}\fi}
\def\lsim{\ifmmode{\mathrel{\mathpalette\@versim<}}
    \else{$\mathrel{\mathpalette\@versim<}$}\fi}
\def\@versim#1#2{\lower 2.9truept \vbox{\baselineskip 0pt \lineskip
    0.5truept \ialign{$\m@th#1\hfil##\hfil$\crcr#2\crcr\sim\crcr}}}
\catcode`\@=12
\documentclass{rspublic}

\usepackage{psfig}

\begin{document}

\title[Active Galaxies and Radiative Heating]{Active Galaxies 
and Radiative Heating}

\author[J. P. Ostriker and L. Ciotti]{Jeremiah P. Ostriker$^1$ and 
Luca Ciotti$^2$}

\affiliation{$^1$Institute of Astronomy, Cambridge, UK\\
$^2$Dept. of Astronomy, University of Bologna, Italy}

\label{firstpage}

\maketitle

\begin{abstract}{Galaxies: active - cooling flows - evolution}
There is abundant evidence that heating processes in the central
regions of elliptical galaxies have both prevented large-scale cooling
flows and assisted in the expulsion of metal rich gas. We now know
that each such spheroidal system harbors in its core a massive black
hole weighing approximately 0.13\% of the mass in stars and also know
that energy was emitted by each of these black holes with an
efficiency exceeding 10\% of its rest mass. Since, if only 0.5\% of
that radiant energy were intercepted by the ambient gas, its thermal
state would be drastically altered, it is worth examining in detail
the interaction between the out-flowing radiation and the equilibrium
or inflowing gas. On the basis of detailed hydrodynamic computations
we find that relaxation oscillations are to be expected with the
radiative feedback quite capable of regulating both the growth of the
central black hole and also the density and thermal state of the gas
in the galaxy.  Mechanical input of energy by jets may assist or
dominate over these radiative effects.

We propose specific observational tests to identify systems which have
experienced strong bursts of radiative heating from their central
black holes.
\end{abstract}

\section{Introduction}

There are many properties of elliptical galaxies and spiral bulges
(systems which are, at a given luminosity, essentially identical) that
have been difficult to understand. These are the observed regularities
with regard to metallicity, gas content, etc., which have largely been
accepted as given.  Now, we have discovered (cf. Tremaine \etal\  2002)
an important new fact, the apparently universal presence (and mass) of
the central black holes (`BHs'), that may provide the key to
understanding these other properties.  The masses of the central BHs
are known to be tightly correlated with the velocity dispersions of
the spheroidal components (e.g., Gebhardt \etal\  2000, Merritt \&
Ferrarese 2001, Tremaine \etal\  2002), or, alternatively, with the
luminosity of those components (cf.  Magorrian \etal\  1998, Kormendy
2000).  Since, for each BH the radiant energy emitted in accumulating
the corresponding mass is known via the Soltan argument (1982), and
the efficiency is also known to within $\pm ~30\%$ to be 0.1 (cf. Yu
\& Tremaine 2002), close to the maximum permitted, and finally, we
know the typical emitted spectrum of an AGN (cf. Sazonov, Ostriker \&
Sunyaev 2004), we thus know to moderate precision the detailed
properties of the electromagnetic radiation seen by a given atom at
some place in the galaxy.  Of course it is possible that the black
hole now seen in a given galaxy has been added at a late time (or a
significant fraction of its mass has been accreted via in-falling
BHs), but that does not change the overall situation significantly,
since the in-falling BH (given the observed scaling relations) was
central to another scaled model of the currently observed spheroidal
system (see also Ciotti \& van Albada 2001 for a discussion of this 
last point).  
The purpose of this presentation is to show that this
radiation will, via standard physical processes, tend to heat and
expel the metal-rich gas (`feedback' in currently fashionable jargon)
in a way that will match the observed properties of elliptical
galaxies with regard to the amount and metallicity of the gas within
them and the gas expelled by them.  It will also help to explain the
observed shut-off of star formation in these systems at a fairly early
epoch, and the observed limitation on the mass of central black holes.
All of these conclusions are preliminary and suggestive, but the
overall logic seems persuasive.  Before proceeding, we would like to
thank our co-workers, S. Sazonov and R. Sunyaev whose contributions
have been essential (cf. also Sazonov \etal\  2004).

Finally we should note that the radiant output is not the only, nor
even necessarily the dominant mechanism whereby feedback from
accretion onto central black holes can heat gas in elliptical
galaxies.  Binney \& Tabor (1995) have stressed that the mechanical
energy input from radio jets will also provide a significant source of
energy, and much detailed work has been performed to follow up this
suggestion.  It is not obvious that all AGNs produce radio jets,
whereas all do appear to produce high energy radiation.  In any case,
the two mechanisms are complementary, and in this paper we are
exploring only the radiative feedback, which may be supplemented by
(or dominated by in some cases), the mechanical energy input.

\section{Elliptical galaxies and spheroidal bulges:\\ properties and problems} 

There are two essential points to be made.  The first is well-known:
elliptical galaxies and spiral bulges form, to good accuracy, a one
parameter sequence.  While a treatment in terms of two parameters via
the `fundamental plane' is slightly more accurate, a single parameter,
the velocity dispersion of the stars within the half-light radius,
$\sigma_e$, provides an excellent predictor of the optical luminosity,
the color, the metallicity, the half-light radius, and now the mass of
the central black hole.  The papers of Kormendy over the years (e.g.,
Kormendy 1977 et seq.) provide the best introduction to this primal,
galactic fact, which originated with the Faber-Jackson (1976)
relation.  Meisels \& Ostriker (1984) elaborated on this fundamental
relation to include also the effects of environment in helping to
determine some aspects of the disc component seen in many systems.
The second point is that this set of relations is most easily
understood (cf. for example Dekel \& Silk 1986) in terms of a feedback
process that terminates galactic stellar and chemical evolution in a
fashion that correlates with $\sigma_e$, or, alternatively, with the
circular velocity, $v_{\rm cir}$, or with the central potential, $\phi_0$,
all of which are tightly correlated with one another.

In general, elliptical components show a low ratio of gas-to-total
mass, with tabulated values from well studied low red-shift examples
typically showing of order 1\% in gas (cf. Knapp, Turner \& Cunniffe
1985), and most of that is seen in X-rays and close to the virial
temperatures of the systems ($\sim 10^{6.5}\,$K).  Standard evolutionary
calculations of a closed box model in which gas is converted to stars
and constantly recycled with the metallicity increasing each
generation (cf. Ostriker \& Tinsley 1975) predict that the gas-phase
metallicity, $Z_{\rm gas}$, will increase as the gas fraction, $f_{\rm gas}$,
decreases:

\begin{equation}
Z_{\rm gas} = -\alpha\ln(f_{\rm gas}),\quad\hbox{where}\quad
\alpha= {Y R\over 1 -R}\simeq 0.01,
\end{equation}
Here, $Y$ is the `yield' and $R$ is the recycled stellar fraction, and 
the single parameter, $\alpha$ is determined by a fit to the mean
stellar metallicity.  The measured metallicities of the gas
\begin{equation}
Z_{\rm gas,obs}\leq 0.03\quad\Rightarrow\quad f_{\rm gas}\geq 0.05.
\end{equation}
Since in fact
$f_{\rm gas} < 0.05$, a significant amount of it has been expelled from
the galaxies in question.  This probably occurred when the ratio of
gas-to-stars was of order 10\%, and, since the gas phase metallicity
is always close to twice the mean stellar metallicity, the total
metals in the expelled gas was probably at least 20\% of the metals in
the stars.  This argument is consistent with the fact that within
clusters the total amount of metals in the intra-cluster gas
(presumably expelled from the galaxies) is of the same order as
the total amount locked up in stars (Renzini \etal\  1993).
Thus the inferred evolution of the systems is that the initial
complement of gas, plus any infalling and recycled gas from dying
stars, was incorporated into successive generations of stars, with an
approximately constant ratio of growing BH mass to stellar mass (at
0.13\%) until the gas fraction was reduced to some level such that
cooling could not keep up with heating (to lowest order the first is
quadratic in the density and the second is linear, so, all other
things being equal, a transition will occur at some specific density
determined by the radiation field).  Then, following the classic
calculations of Dalgarno \& McCray (1972), the gas undergoes a
transition (at constant pressure) to a higher temperature, lower
density phase. Evidence for the constancy of the ratio of the
accretion rate to the star formation rate can be found in Haiman,
Ciotti \& Ostriker (2004) and Heckman \etal\  (2004).  A detailed
argument for the transition to the high temperature solution at
$M_{\rm gas}/M_*\simeq 0.01$ can be found in Sazonov \etal\  (2004).  A
considerable amount of gas will be expelled at this time and,
subsequently, both star formation and black hole accretion will
proceed at a much lower level, as determined by the availability of
additional gas and the accretion and infall possible in the high
temperature phase.

A simple quantitative estimate is possible by noting that, if a mass
roughly equal to 10\% of the stellar mass was expelled, and the escape
velocity is approximately $700\kms$, then the energy required to do
this, in units of the total stellar mass, can be expressed as a
feedback efficiency, $\epsilon_{*fb}$:
\begin{equation}
{\Delta E \over M_* c^2}\equiv \epsilon_{*fb} =10^{-6.3}.
\end{equation}
If, instead we consider that the source of the energy may have been
the central BH, having a mass equal to $\Mbh=10^{-2.9}\Mstar$, then we
can rewrite (2.3) as
\begin{equation}
{\Delta E \over \Mbh c^2}\equiv \epsilon_{\rm BHfb} =10^{-3.4}.
\end{equation}

Now the BH has emitted radiant energy with an efficiency of
$10^{-1.1}$, so only a fraction of the emitted energy $10^{-2.3}$ is
required to expel the gas.  In sum, if approximately 0.5\% of the
radiant emission of the central AGN can couple to the gas, this will
suffice to expel the required amount.  We will show later that this is
exactly what is expected to occur, given the emitted spectrum and
normal atomic processes.  But first, let us note an argument,
additional to the two mentioned so far (which depend on the observed
metallicity of the galaxy and the amount of cluster gas respectively),
that indicates the likely importance of central feedback.

There is a well-known `cooling flow problem' in clusters of galaxies:
the time for a significant fraction of the centrally located gas to
cool ({\it via} the observed radiative output) and to collapse is
small compared to the age of these systems, smaller in comparison to
the Hubble time in ellipticals than in clusters.  A source of energy
must be found that can convincingly produce a balance between energy
loss and gain (Fabian 2004). In ellipticals the cooling times are
typically $\gsim 10^6$ years (Ciotti \etal\  1991), significantly
shorter than the billion-year timescale for intra-cluster gas.  Once
again, the solution must be a source of energy, feedback, from the
stellar or central BH energy sources.  The energy requirements are
comparable to those noted for expulsion of the gas earlier noted, and
the observed rate of supernovae in such systems is quite capable of
supplying the needed energy.  However, it is not likely that SNIa
provide the entire solution, because the distribution of the energy
input (parallel to the observed light profiles) is not nearly
concentrated enough to balance the observed gas cooling rates (which,
since they scale as $\rho^2$, are required to be very large in the
very central regions).  Moreover, the rate of SNIa is independent of
the current thermal state of the X-ray emitting plasma, so SN heating
cannot act as a self-regulating mechanism.  Thus, a concentrated
feedback source is well designed for this purpose, and the central BH
is the natural candidate, by its mass and by its location through a
combination of mechanical and radiative feedback mechanisms
(cf. Binney \& Tabor 1995, Ciotti \& Ostriker 1997, for a review see
Mathews \& Brighenti 2003).

Let us now examine the situation in somewhat greater detail.  The
`standard' cooling flow model, even though observationally motivated
and rich with testable predictions, nonetheless presents a few major
unsolved problems when applied to elliptical galaxies. Perhaps the
most severe is the fate of the supposedly cooling material. In fact,
in elliptical galaxies the mass return rate from the (passively)
evolving stellar population is well known, and is of the order of
$1.5\times 10^{-11}\Lb$ solar masses per year, where $\Lb$ is the
galaxy blue luminosity in solar units (Ciotti \etal\  1991).  Thus, a
long lived cooling flow would accumulate mass in a central BH with
mass substantially exceeding that currently observed.  Alternative
forms of cold mass disposal, such as distributed mass drop-out/star
formation, have been proved not viable solutions (e.g., Binney 2001).

A solution to this problem was proposed by D'Ercole \etal\  (1989) and
Ciotti \etal\  (1991), by considering the effect of SNIa heating of the
galactic gas, and exploring the time evolution of gas flows by using
hydrodynamical numerical simulations. Subsequent, more realistic
galaxy models, also with the up-to date rates of SNIa (as derived by
direct counts from optical observations) were explored by Pellegrini
\& Ciotti (1998). Basically, it was found that SN input sufficed for
low and medium-luminosity elliptical galaxies to produce fast winds,
but the more massive spheroids would still host inflow solutions
similar to cooling flows.  The physical reason for this behaviour is
just the fact that, while the number of SNIa per unit luminosity is
expected to be constant in all ellipticals, the gas binding energy per
unit mass increases with galaxy luminosity.  Thus, feedback is still
required for medium-large ellipticals or else they will experience
significant mass accretion onto their central BH, producing masses
much larger than those observed (Binney \& Tabor 1995).

\begin{figure}
\centerline{\psfig{file=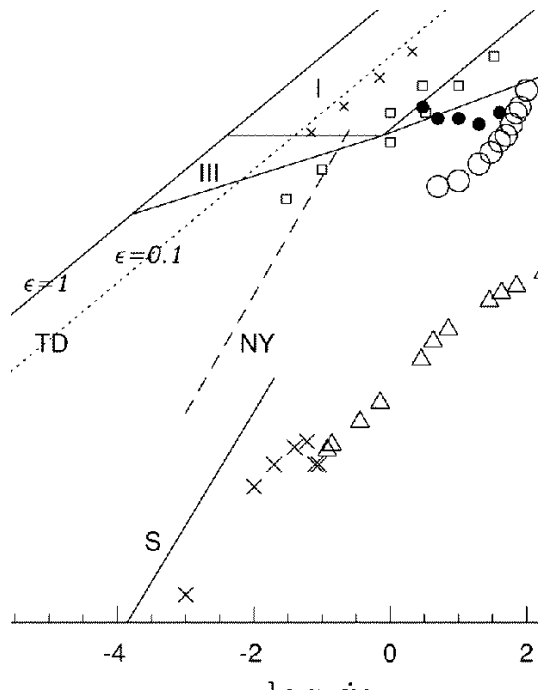,width=7cm,angle=0}}
\caption{Spherical and disk accretion flow solutions in the ($l,\dot
m$)-plane ($l\equiv L/\Ledd$, $\dot m\equiv \dot M c^2/\Ledd$). Solid
line S represents low-$\dot m$ spherical accretion solution of
Shapiro (1973), and large crosses are those of Park (1990).
Low-temperature solutions of high-$\dot m$ spherical accretion are
triangles (Nobili \etal\  1991), and high-temperature solutions are
large open circles (Park 1990). Accretion disk solutions appear as
the dotted line TD for thin disk, the dot-dashed line SD for slim disk and
the dashed line NY for ADAF (Narayan \& Yi 1995). Spherical accretion
solutions with electron-positron pairs are small filled circles (Park
\& Ostriker 1989), those with the magnetic field reconnection small
squares, and those with preheating shocks small crosses (Chang \&
Ostriker 1985) (from Park \& Ostriker 2001).}
\end{figure}

\section{Quasi-spherical accretion on MBHs: empirical and theoretical aspects}

The classical work on spherical accretion to a BH (e.g., Shapiro 1973)
assumed Bondi accretion (1952) at large radii and looked for steady
state solutions having a mass flow rate independent of radius.  
Radiant losses from the accreting gas were included but
energy gained when fluid elements absorb
some of the radiation emitted by interior, energy losing regions was neglected.
Modifications to the classic work have either abandoned spherical
symmetry, the assumption of a mass flow rate independent of radius,
the assumption of a steady state or have allowed for energy gains as
well as losses.  While the inclusion of rotation certainly requires
one to abandon spherical symmetry, the `Polish Doughnut' (Jaroszynski,
Abramowicz \& Paczynski 1980), and other solutions which are
physically thick are not, in essentials, very different from the
corresponding spherical solutions.  But the ADAF (Narayan \& Yi 1994)
and CDAF (e.g., Quataert \& Gruzinov 2000) solutions, which are not
mass conservative, are substantially different from the classical
fixed mass flow solutions.  The revisions required when one allows for
radiant energy gains (`feedback' again) are still more substantial and
typically preclude steady flow except in cases of very low luminosity
(compared to the Eddington value).  As an example of combining two of
the modifications, Park \& Ostriker (2001) showed that the standard
ADAF solution becomes physically inconsistent in the polar regions,
because the slow infall near the pole allows the optically thin gas
enough time to be heated by the radiation emitted from interior
regions sufficiently to reverse the flow.  Thus ADAF and CDAF flows
would tend to develop evacuated, conical regions with intermittent
winds likely emitted as unsteady jets (cf. Ryu \etal\  1995 and Das
\etal\  2001 for analogous solutions).

Turning to quasi-spherical flows, Park \& Ostriker (1998),
summarizing the extant literature, noted that hot flows tend to become
intermittent, due to preheating of the inflowing gas when the flow
rate is within two orders of magnitude of the Eddington limit
(cf. figure 1).  More generally, the nature of the flow can be
characterized by two dimensionless numbers: the luminosity in units of
the Eddington luminosity and the mass flow in the same units ($l, \dot
m$).  For low values of both quantities, the gas density is everywhere
low, and both emission and absorption of radiation are negligible.
The flow is essentially adiabatic.  There exist self-consistent flows
with high mass flow rates and low luminosity; in these the gas cools
at large distance from the black hole, and once on a low adiabat, it
radiates inefficiently only a small fraction of its total kinetic
energy but cools efficiently, losing a large fraction of its
(relatively small) thermal energy. So the flow stays cool all the way
to the horizon.  There would be not much of an observational signature
for such flows, so they may be more important than is now known in
building up the mass of the central black holes.

For relatively high values of the mass flow, there also exist high
temperature solutions.  In these cases, radiation from the inflowing
dense gas at small distances from the black hole is intense enough to
heat the gas at large radii at a rate greater than it cools, so it
stays on a high adiabat and hence, self-consistently, when it arrives
near the black hole it is hot and dense and capable of emitting the
radiation needed to maintain the hot flow.  Typically, these solutions
are possible (and in fact necessary) whenever the luminosity is within
two orders of magnitude of the Eddington value. Time dependent
computations typically find that these solutions are unstable, since
downward fluctuations in the luminosity lead to less preheating and
thus lower luminosity -- and {\it vice versa}.  A rapid flaring
behaviour ensues.  The overall average luminosity is set by the infall
rate (roughly determined by the Bondi formulation at the accretion
radius which is outside the flaring zone), so the luminosity
cycles between the Eddington value and a much lower value and results
in a small duty cycle.  On a longer timescale there can be heating of
gas outside the accretion radius that will lower the accretion rate
significantly and essentially shut off the accretion process until the
gas in the bulk of the galaxy cools sufficiently to again begin
accretion.  Figures 2 and 3 illustrate these two types of inflow of in
a case with parameters chosen to represent typical accretion of a
massive black hole in an elliptical galaxy (Ciotti \& Ostriker 1997).

\begin{figure}
\centerline{\psfig{file=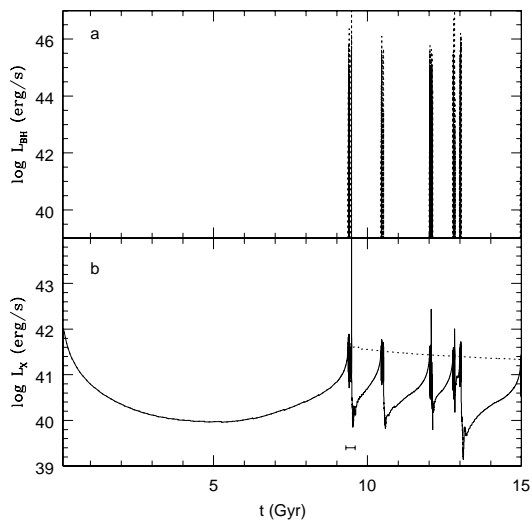,width=7cm,angle=0}}
\caption{{\it Panel a}: the time evolution of $\Lbh$ (bolometric)
emitted at the galaxy centre. {\it Panel b}: the time evolution of
$\Lx$ for the model with $\epsilon = 0.1$ (solid line), and that of
the same model with $\epsilon =0$ (cooling flow -- dotted line).  $\Lx$
is calculated inside the galaxy truncation radius and in the range 0.5
$4.5\,$KeV. Time interval in horizontal error bar is expanded in figure 3
(from Ciotti \& Ostriker 1997).}
\end{figure}

\begin{figure}
\centerline{\psfig{file=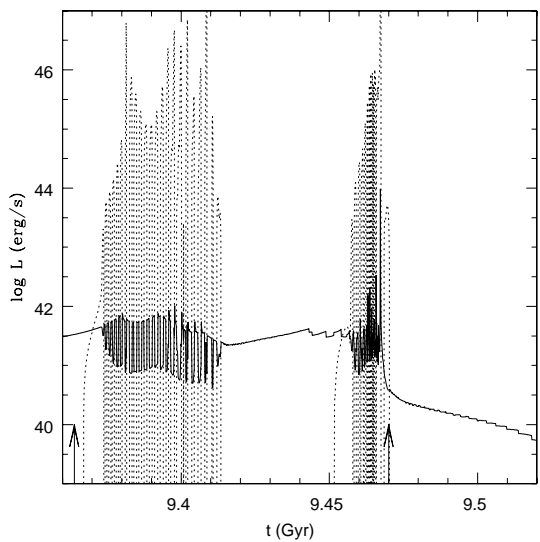,width=7cm,angle=0}}
\caption{Time expansion of the first burst shown in figure 2.  The
solid line is $\Lx$, the dotted line $\Lbh$.  The temporal
sub-structure of the burst is apparent, and the quasar-like luminosity
$10^{45} <\Lbh /{\rm erg\,s^{-1}} <10^{47}$ is seen during bursts. Arrows
mark epochs of ``before'' and ``during'' bursts referred to in text
(from Ciotti \& Ostriker 1997).}
\end{figure}

\section{Results: hydrodynamic simulations of an accreting black hole in an 
elliptical galaxy}

Figure 2 shows the evolution of the luminosity computed over a Hubble
time, while figure 3 shows a close-up of the first series of bursts,
approximately 2.5 Gyr from the onset.  Overall the duty cycle computed
(the fraction of the time that the AGN is in the `on' state and
radiating near the Eddington limit) is 0.006, quite low and consistent
with the fraction of low redshift elliptical galaxies which are seen
in the active, quasar state.  Additionally, the distribution of X-ray
luminosities seen above can be compared to that observed in a local
sample of elliptical galaxies and is in good agreement.  In contrast,
if one did not allow for the feedback, not only would the duty cycle
be far higher (as nothing significantly would be impeding the
accretion), but the typical luminosities would be higher than those
seen.  Next we turn to the luminosity distribution (taken at random
time intervals) of the X-ray emission from the hot gas in the galaxy.

Figure 4 shows that the agreement with the observed distribution of $L_X$ is
good, much better than if we ignore feedback (`cooling flow' model).
Finally, the resultant black hole masses are computed to be in the
observed range.  In the absence of the feedback, the masses grow to
values ($\sim 10^{11}\Msun$) far higher than observed anywhere.

\begin{figure}
\centerline{\psfig{file=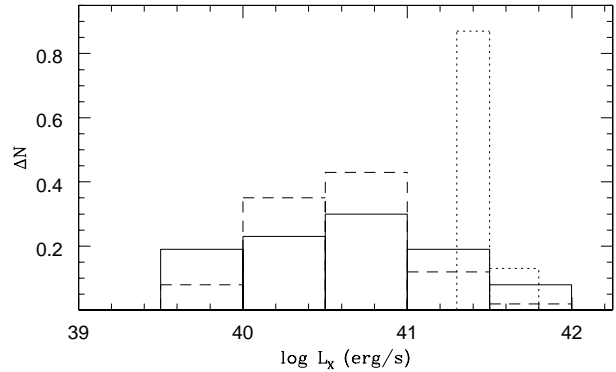,width=8cm,angle=0}}
\caption{The statistical distribution of $\Lx$ for observed galaxies
(solid) in the range $10.4<\log\Lb/\Lsun <10.8$ derived from figure 1
of Ciotti \etal\  (1991).  The dashed histogram represents the time
distribution of $\Lx$ for the model in Fig. 2 from $9$ to $15\,$Gyr,
while the dotted histogram shows the cooling flow ($\epsilon =0$)
model; clearly the bursting model provides a better fit to the
observed distribution of $\Lx$ (from Ciotti \& Ostriker 1997).}
\end{figure}

In summary, if one allows the radiation emitted from the accreting
black hole to interact with and heat the gas in the galaxy from which
it is accreting, one solves the cooling flow problem in elliptical
galaxies, and the feedback produces systems that typically look like
normal ellipticals containing hot gas. They sometimes look like incipient
cooling flows and rarely but importantly look like quasars.  In the
works quoted (Ciotti \& Ostriker 1997, 2001), there was a major
uncertainty in not knowing with any accuracy the typical QSO spectrum,
in particular the high energy component of that spectrum, which is
most important for heating the ambient gas.  Thus, a simple broken
power law was adopted for the spectrum with a range of possible values
of the Compton temperature -- from $10^{7.2}\,$K to $10^{9.5}\,$K -- with
most of the emphasis of the paper being on the higher temperatures.
Subsequent work by Sazonov, Ostriker \& Sunyaev (2004), which
carefully assessed the full range of observational data from AGNs,
concluded that the typical radiation temperature was narrowly bounded
to values near $10^{7.3}\,$K, at the lower end of the range adopted by
Ciotti and Ostriker, thus reducing the typical heating rate.  However
there is a rather large compensating effect also not included by
Ciotti and Ostriker: gas heated by radiation with a characteristic
temperature near $10^7\,$K is heated far more effectively by absorption
in the atomic lines of the abundant metal species than by the Compton
process.  With these two modifications, Ciotti and Ostriker have
recomputed the evolution of an accreting black hole in elliptical
galaxies.  The detailed results will be presented in forthcoming
papers and only a brief summary will be given here.  Two types of
calculation were made.  In one, labeled the `Toy Model', we have
allowed for a large number of relevant physical processes but treated
the mathematics of accretion casually, by adopting a two-zone model
and assuming Bondi-like accretion.  In the second, labeled
`Hydrodynamic', we do a careful job with regard to spatially and
temporally resolving the hydrodynamic flow problem (shocks, radiative
transfer and other details computed) but neglecting physical
complications such as star formation, cosmological inflow etc.

\subsection{The `Toy Model' approach}

The adopted equations for the mass budget in the Toy Model are given
below, with subsequent figures presenting some of the salient results.
Their full description is given in Sazonov \etal\  (2004), together
with the energy equations (not shown here).  The gas budget for the
galaxy (equation [4.1]) allows for cosmological infall (equation
[4.2]), star formation (equation [4.3]), recycled gas from dying stars
(equation [4.4]), accretion onto the central black hole(s) (equations
[4.6]-[4.7]), and possible galactic superwinds (equation [4.8]).  Star
formation follows the prescriptions made familiar by the semi-analytic
models.  Seed black holes are assumed to be produced by the collapse
of massive stars following normal stellar evolution.  Dynamical
friction brings them to the center of the system where they are
assumed to merge.

\begin{equation}
\DMgas = \DMinf - \DMstar +\DMrec - \DMbh -\DMesc.
\end{equation}
\begin{equation}
\DMinf = {\Mgal\over\tinf}\exp\left(-{t\over\tinf}\right),\quad
(\tinf\simeq 2.5\,\hbox{Gyr}).
\end{equation}
\begin{equation}
\DMstar = \DMstar^+ - \DMrec,
\end{equation}
\begin{equation}
\DMrec = \int_0^t \DMstar^+ (t')\Wstar (t-t')\; dt',
\end{equation}
where
\begin{equation}
\DMstar^+ ={\alstar \Mgas \over\max (\tdyn , \tcool)},
\quad (\alstar\simeq 0.1\div 0.3),
\end{equation}
and the function $\Wstar$ is given in Sazonov \etal\  (2004).
\begin{equation}
\DMbh = \DMbhac + \bhstar\DMstar^+, \quad (\bhstar\simeq 1.5\times 10^{-4}),
\end{equation}
where
\begin{equation}
\DMbhac = \min (\fed\DMedd , \DMbon),
\end{equation}
and $\DMbon$  is the Bondi accretion rate. Finally
\begin{equation}
\DMesc ={\Mgas\over\tesc},\quad T \geq \eta_{\rm esc}\Tvir,
\end{equation}
and $\DMesc=0$ otherwise. 

Overall, there are two phases, one in which star formation is the
dominant process and the growing black holes have a negligible effect,
and then later, after the gas reservoir is significantly depleted, the
by now massive black holes heat and eject much of the remaining gas in
co-operation with the remaining supernova induced heating.  The first
phase would be identified observationally with the Lyman Break
Galaxies.  Roughly 3\% of these (Steidel \etal\  2002) show central
AGNs, thus identifying the duty cycle at this time with roughly this
level.  Such systems are in the stage of rapid star formation and show
significant wind activity.  The later, high gas temperature phase
would be identified with normal, local ellipticals which contain
little gas, have low rates of star formation and have a duty cycle
(fraction of time during which they appear as AGNs) of roughly 0.1\%.

Thus, in equation (4.7) the duty cycle factor is known empirically to
be $\fed \simeq 0.03$ in the cold phase (i.e., $\tcool/\tdyn <1$),
while $\fed \simeq 0.001$ in the hot phase (i.e., $\tcool/\tdyn >1$).

\begin{figure}
\centerline{\psfig{file=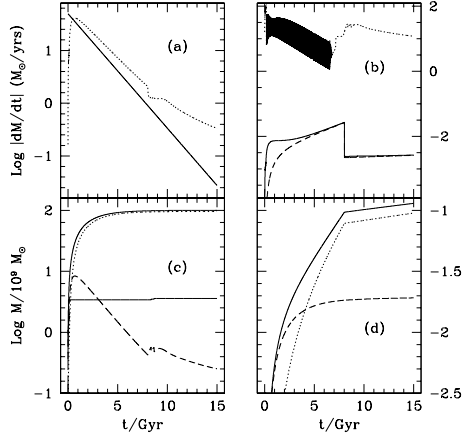,width=6cm,angle=0}}
\caption{Results for a galaxy with an effective radius of 4 kpc,
circular velocity of $400\kms$, $\epsilon = 0.1$.  The duty cycle is
0.01 in the `cold phase' and 0.001 in the `hot phase'.  {\it Panel a}:
mass infall rate (solid line) and stellar mass formation rate (dotted
line).  {\it Panel b}: total BH accretion rate (solid line), Bondi
accretion rate (dotted line), Eddington accretion rate (reduced by the
duty-cycle factor $\fed$, dashed line).  {\it Panel c}: total infall
mass (solid line), total stellar galaxy mass (dotted line), total
galaxy gas mass (short-dashed line); the nearly horizontal line is the
escaped gas mass.  {\it Panel d}: total BH mass (solid line), total
mass gaseously accreted (dotted line), BH mass originated from stellar
remnants (dashed line).}
\end{figure}

\begin{figure}
\centerline{\psfig{file=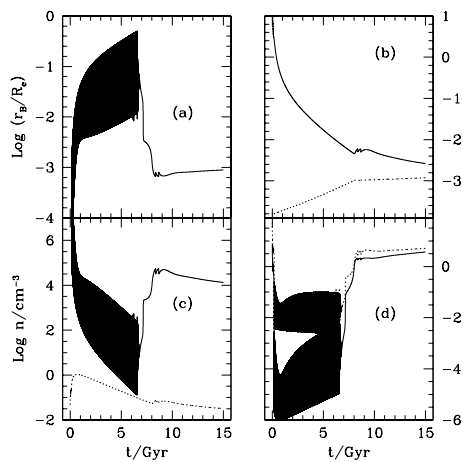,width=6cm,angle=0}}
\caption{{\it Panel a}: time evolution of the Bondi radius.  {\it
Panel b}: logarithm (base 10) of the ratio between gas mass to stellar
mass (solid line) and of the ratio between BH mass to stellar mass
(`Magorrian relation', dotted line).  {\it Panel c}: gas density at
the Bondi radius (solid line) and mean gas density (dotted line).
{\it Panel d}: logarithm (base 10) of the cooling time (solid line)
and heating time (dotted line) measured in terms of the dynamical
time.}
\end{figure}

\begin{figure}
\centerline{\psfig{file=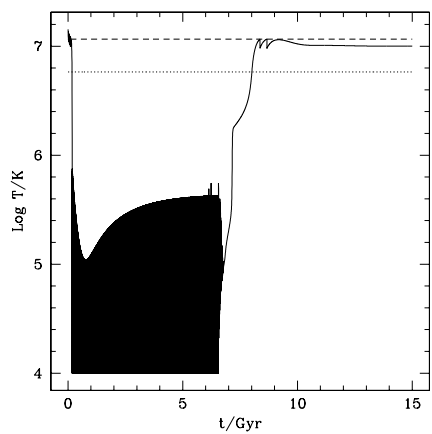,width=6cm,angle=0}}
\caption{Time evolution of the model gas temperature (solid line).
The model virial temperature is represented by the dotted line, which
the dashed line represents the `escape' temperature (here assumed
$2\Tvir$).  An identical galaxy model with circular velocity of $300\kms$
would present strong degassing in the hot phase.}
\end{figure}

We see, as expected, flaring during the cold flow stage of evolution,
when star formation dominates, and then, when gas is depleted to a
level of about 1\% of the stellar mass, a transition to the hot
solution, some outgassing of the galaxy and a decrease in the rate of
growth of the black hole.  Overall, in the presented model about 80\%
of the growth of the central black holes is from accretion and about
20\% from the merging of stellar seed black holes.

\subsection{The hydrodynamic approach}

Next we turn to the results of the detailed hydrodynamic calculations.
Figure 8 shows the evolution of the gas mass and black hole mass, for
a model similar to the Reference Model in Ciotti \& Ostriker (2001),
calculated with the new heating function and the Compton temperature
as derived by Sazonov \etal\  (2004).

\begin{figure}
\centerline{\psfig{file=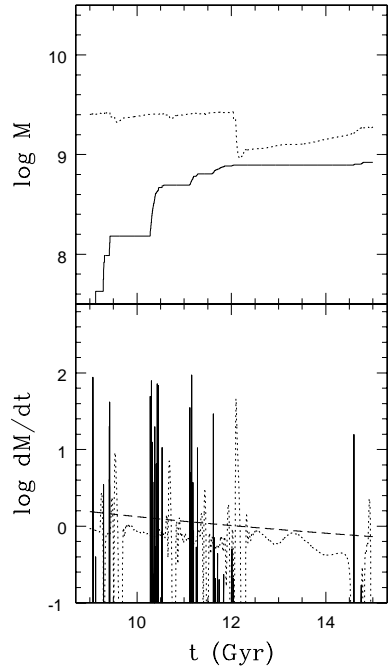,width=5cm,angle=0}}
\caption{The mass budget evolution of a model similar to the
Reference Model of Ciotti \& Ostriker (2001), with Compton
temperature reduced to $2\times 10^7\,$K but with contribution from
line heating. {\it Upper panel}: total galaxy gas mass (dotted line)
and accreted mass on the central black hole (solid line).  {\it Lower
panel}: mass return rate from the evolving stellar population (dashed
line), galactic wind mass loss rate (dotted line), and mass accretion
rate on the central black hole.}
\end{figure}

The left panel of figure 9 shows the temperature and density in the central
regions of the model: note how the QSO bursts heat the central gas, causing
the density to drop; note also the Compton temperature floor. The right panel
shows the time evolution of the luminosity of the central black hole and of
the coronal, X-ray emitting gaseous atmosphere of the galaxy. Note how the
QSO luminosity is grouped in highly structured bursts. In particular, the
X-ray luminosity of the galactic ISM falls in the range of commonly observed
in real galaxies, with mean values lower than the expected luminosity for a
standard cooling-flow model (see figure 10).

\begin{figure}
\centerline{\psfig{file=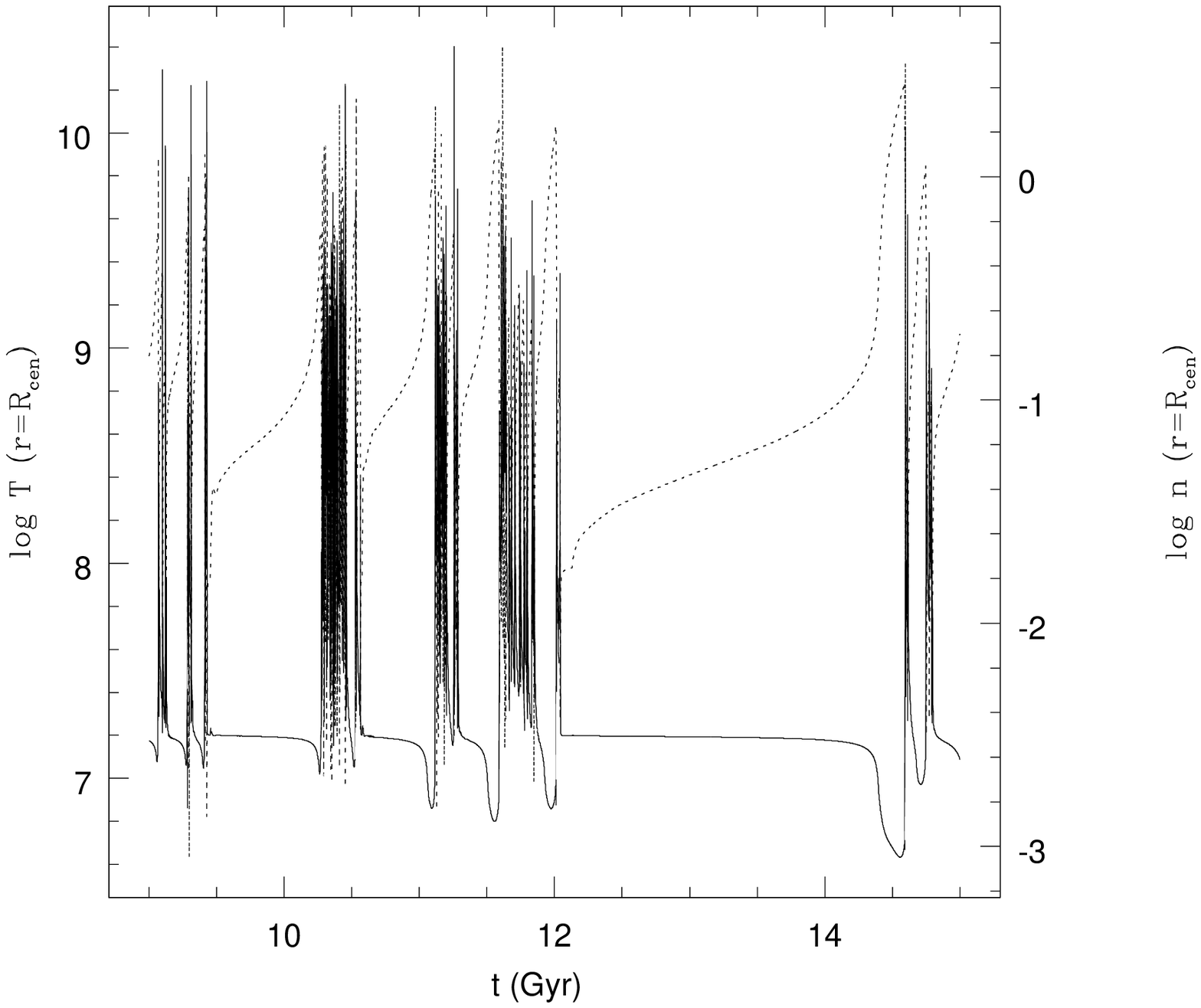,width=6.5cm,angle=0}\quad 
\psfig{file=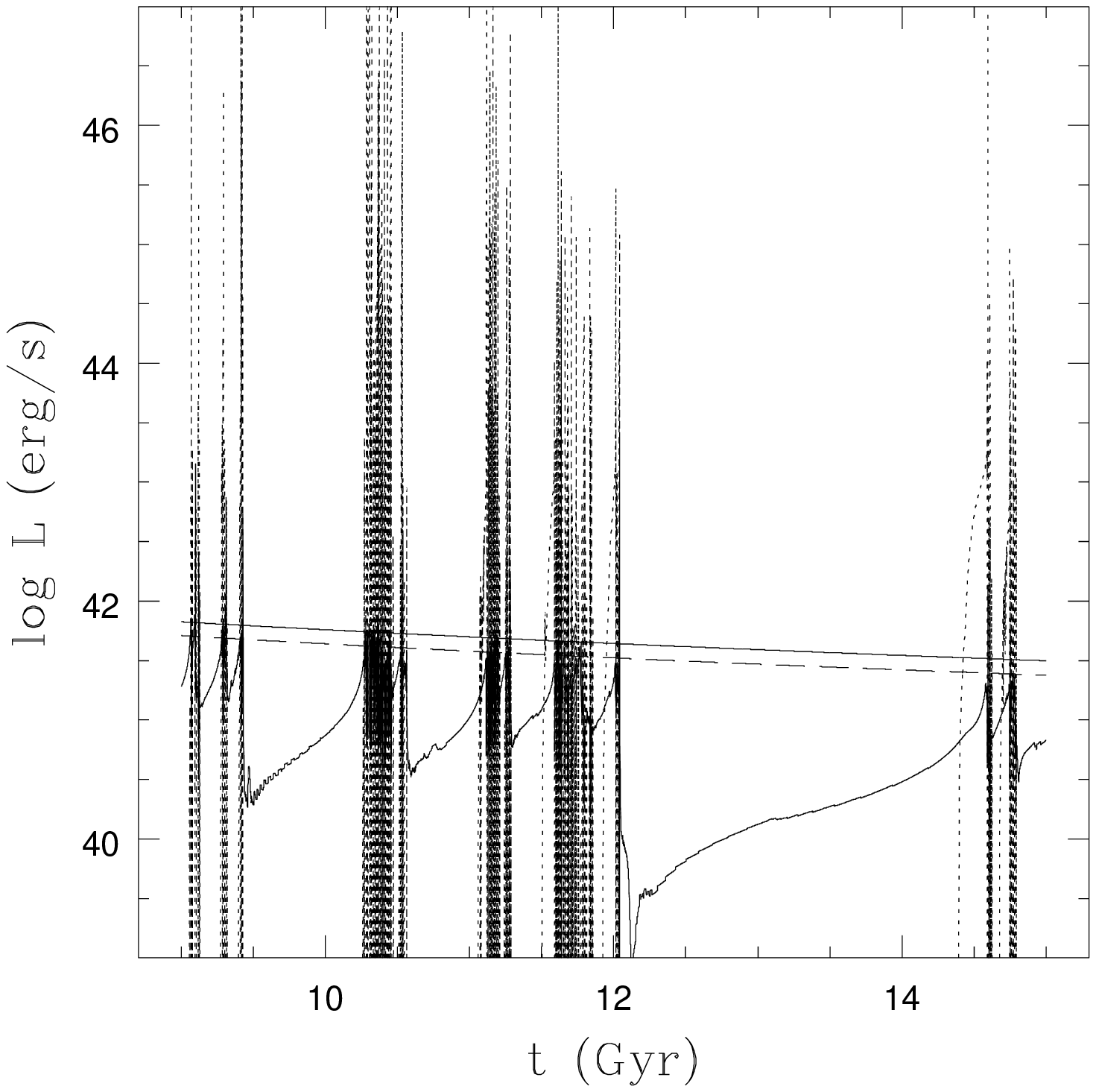,width=5.4cm,angle=0}}
\caption{Time evolution in the model of
figure 8: gas density (left panel, dotted line); temperature
near the black hole ($R_{\rm cen}= 20$ pc) (left panel, solid line);
the accretion luminosity $\Lbh$ (right panel, dotted
line); the galaxy X-ray coronal luminosity $\Lx$ (right panel, solid line).
In the right panel the nearly horizontal solid line
represents the energy per unit time required to steadily extract the
gas lost by the stars from the galaxy potential well. The energy per
unit time provided by SNIa is shown as the dashed line.  Note how the
QSO luminosity is grouped in bursts.}
\end{figure}

\begin{figure}
\centerline{\psfig{file=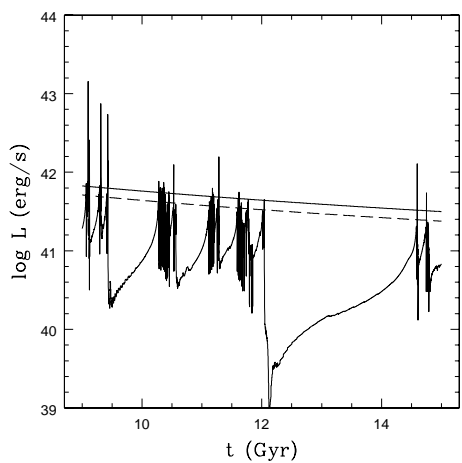,width=6cm,angle=0}}
\caption{Same plot as in figure 9, where the accretion luminosity
has been eliminated. The solid line represents $\Lx$.}
\end{figure}

Interesting aspects of the feedback process can be observed looking at
the same data on expanded temporal scales, as shown in figures
11--12. In particular, it is apparent how the major bursts are
organized in several bursts with shorter and shorter time-scales.

\begin{figure}
\centerline{\psfig{file=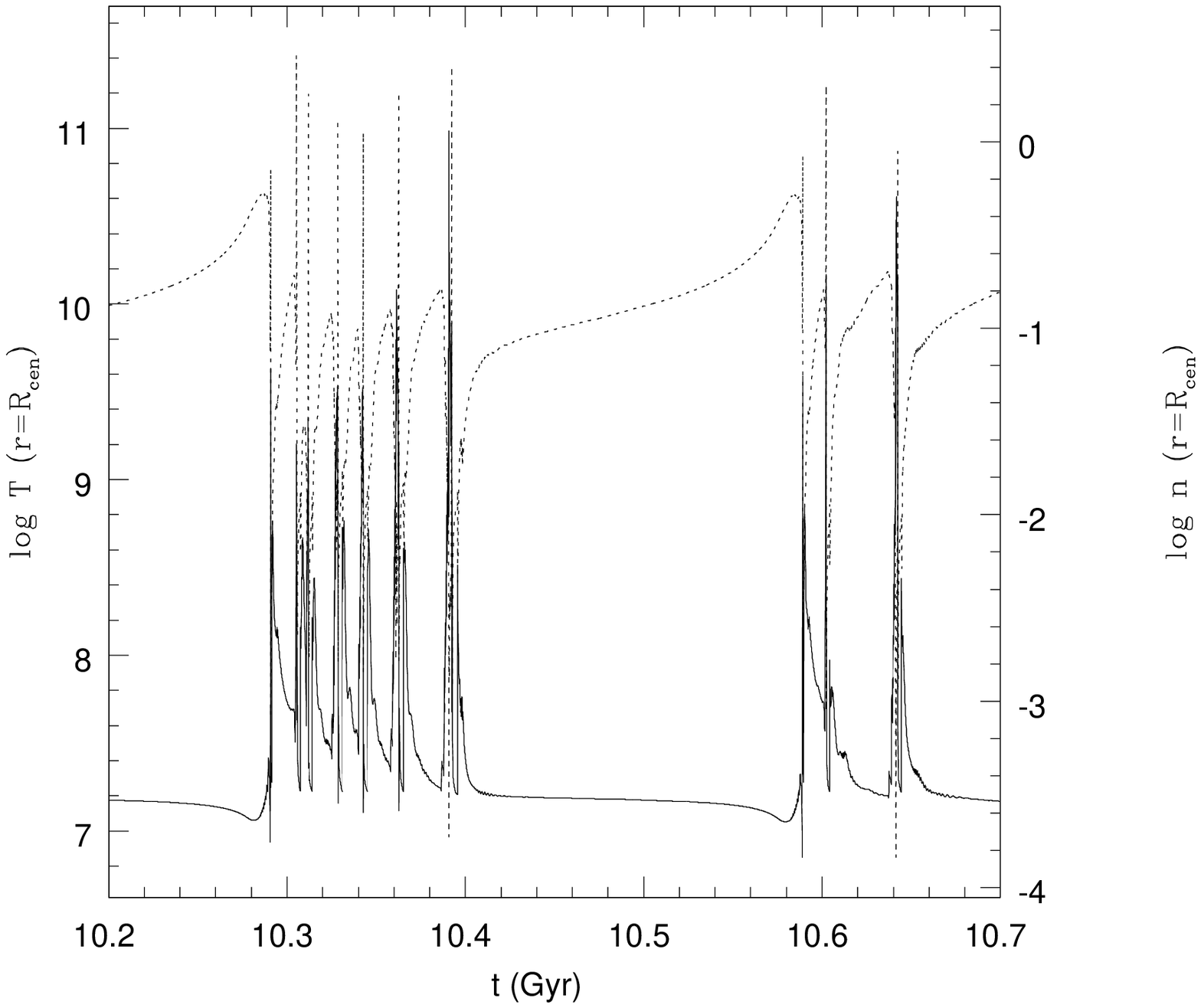,width=6.5cm,angle=0}\quad
\psfig{file=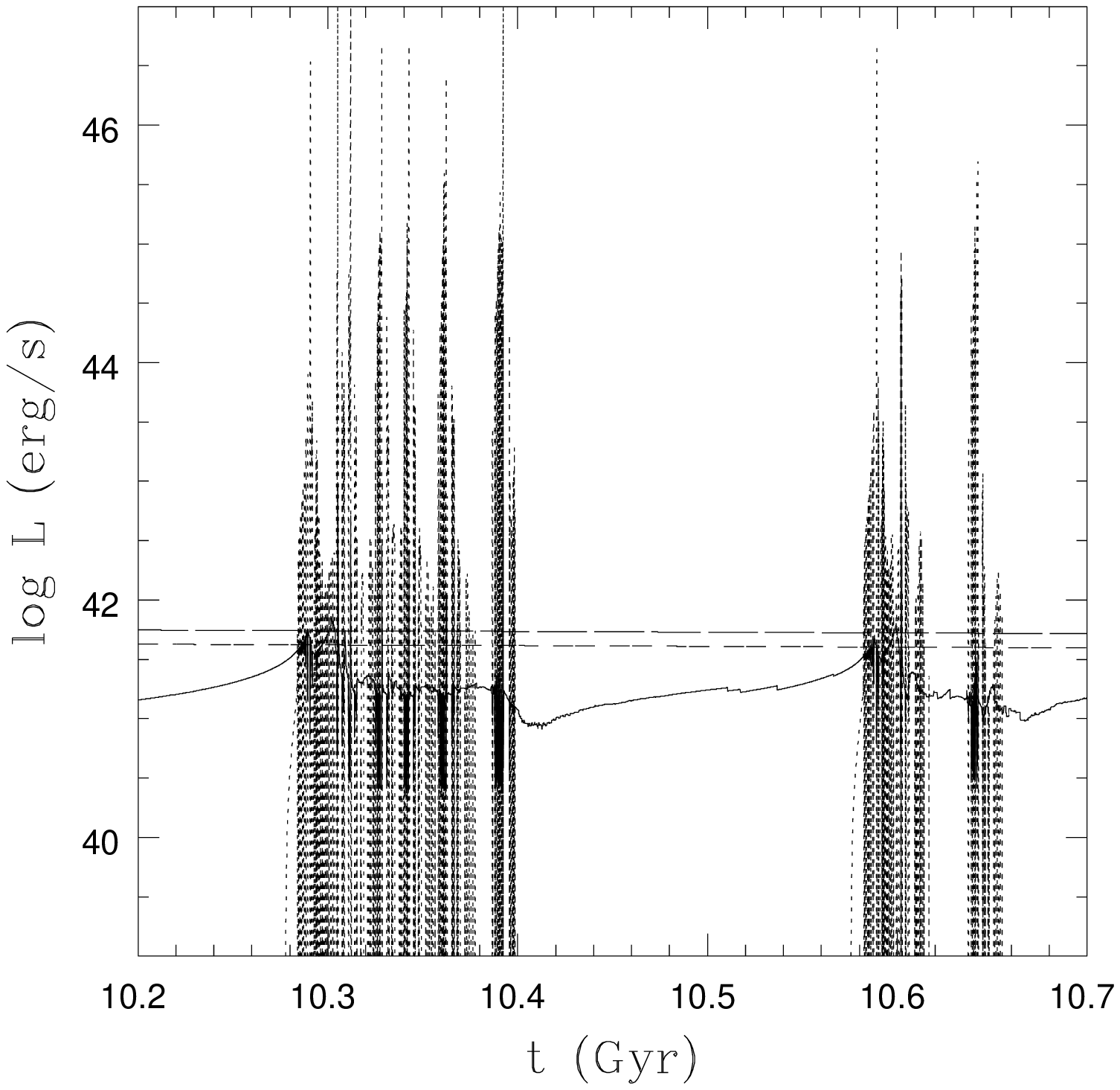,width=5.5cm,angle=0}
}
\caption{Time expansion of the major burst at $t\simeq 10\,$Gyr in
figure 9 with $T$ at left and $L_X$ on right.
Note how the QSO bursts of
$\sim 10^{46}\ergs$ initially heat the gas and increase its
luminosity, then cause expansion and lower X-ray luminosity. After a
major burst significant nuclear activity is suspended until onset of
cooling.
}
\end{figure}

\begin{figure}
\centerline{\psfig{file=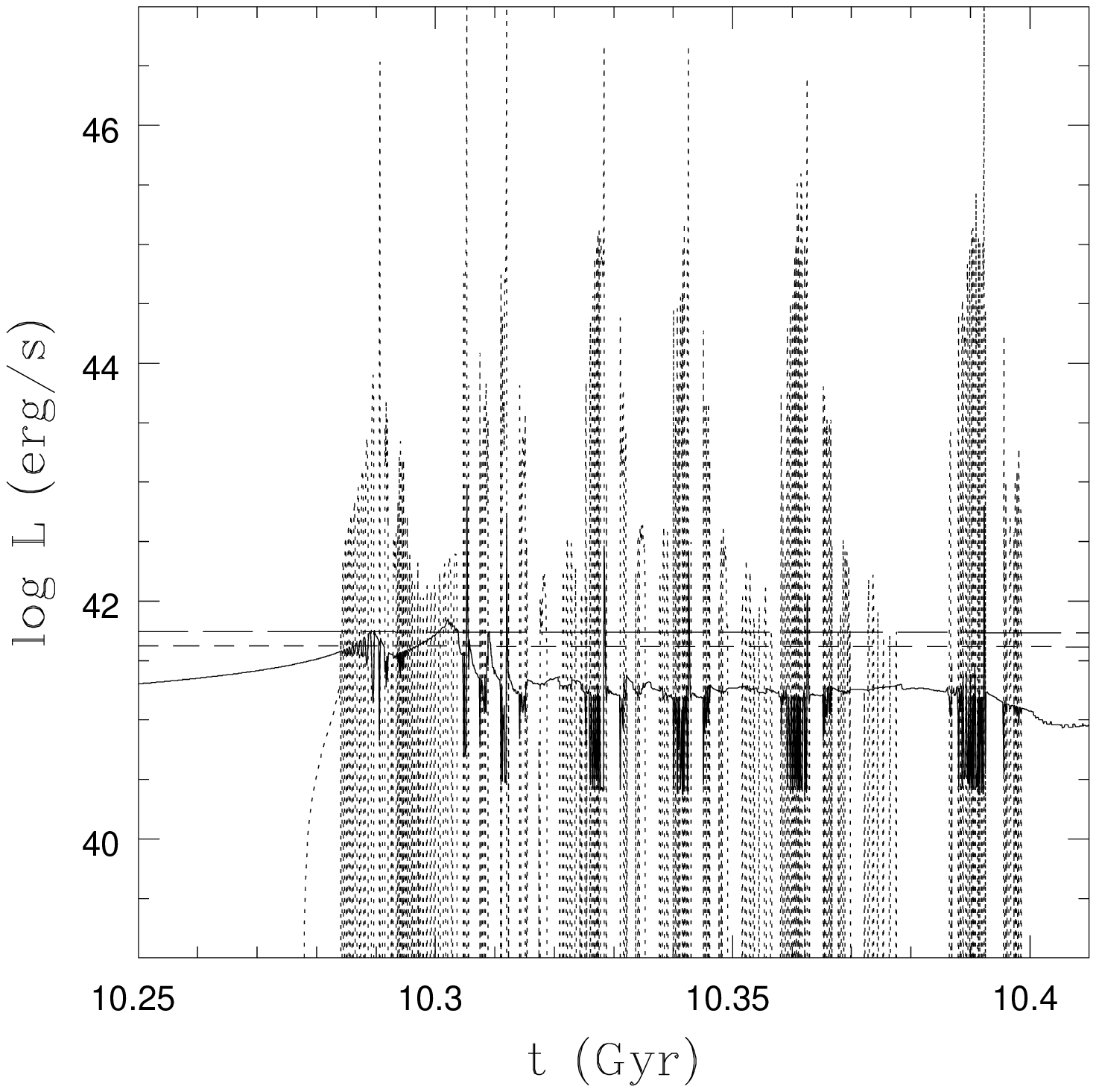,width=6cm,angle=0}\quad
\psfig{file=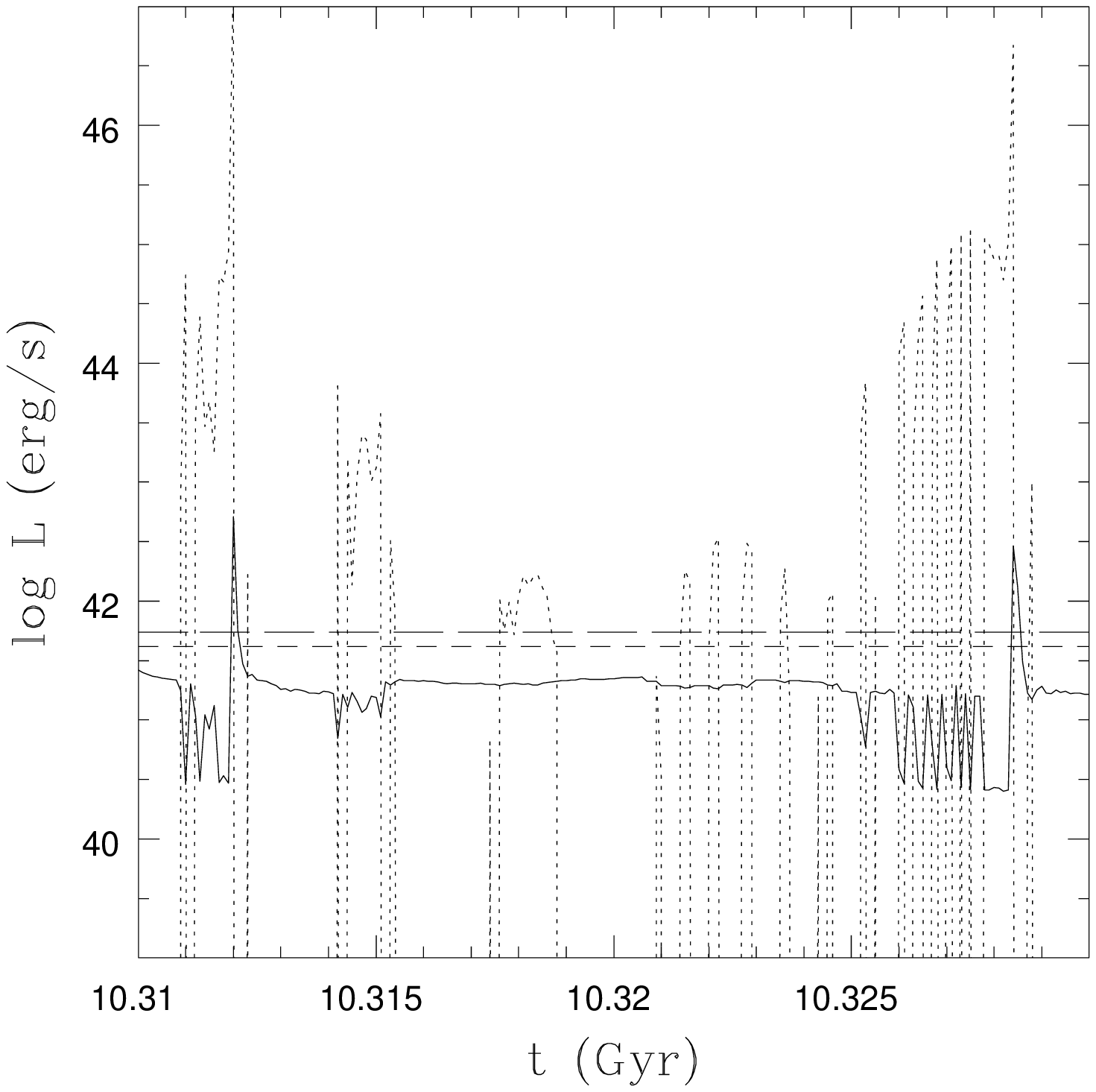,width=6cm,angle=0}}
\caption{Two further temporal zooms to resolve QSO output into
short time-scale bursts. At right the changes in QSO and coronal
X-ray gas luminosity are anti-correlated and time lagged.}
\end{figure}

Thus, the complex flaring behaviour of the accretion shown by the
detailed hydro simulations during the hot phase is superficially
consistent with our knowledge of low redshift QSOs in giant elliptical
galaxies, but there are many details to be computed and compared to
observations. As an example, in figure 13 we show four representative
snapshots of the X-ray surface brightness of the galaxy before and
after a major nuclear outburst.

\begin{figure}
\centerline{\psfig{file=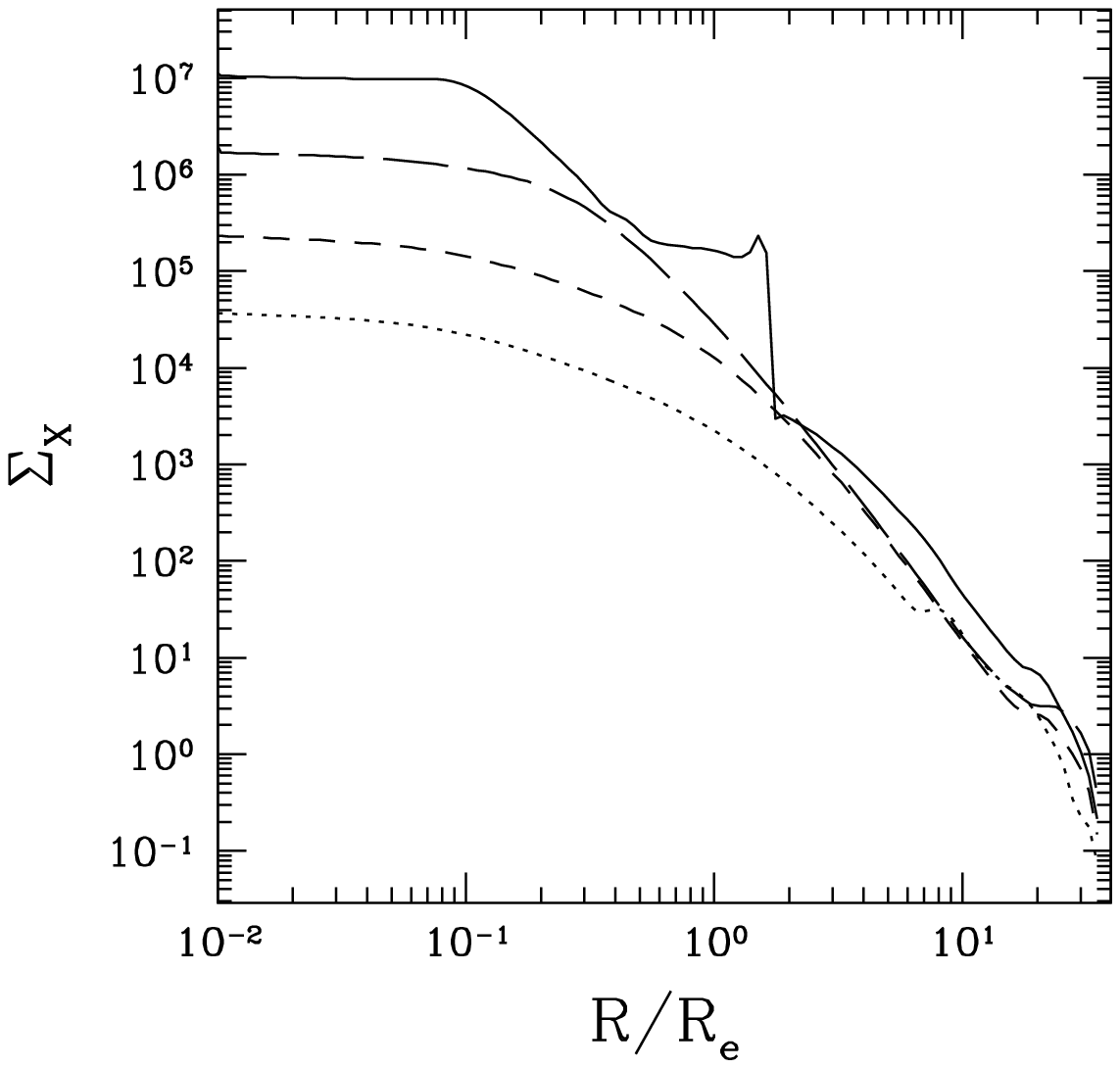,width=6cm,angle=0}}
\caption{X-ray surface brightness radial profiles for the model
galaxy presented. All profiles are normalized to the same (arbitrary)
value. The short-dashed, long-dashed, solid, and dotted lines refer to
the model near the `cooling catastrophe', at the beginning of the
burst, during the burst, and in the following low state,
respectively. Detailed X-ray profiles as a function of energy could
provide the strongest test of the model.}
\end{figure}

\section{Summary and future tests} 

We now know the masses and radiative outputs of the black holes found
in the centers of elliptical galaxies with some reasonable certainty,
and it is clear that the EM output should have had a powerful effect
on the gas resident in these systems.  Detailed, but still uncertain
calculations indicate that relaxation oscillations should be the
normal state as energetic output from the central BH at near the
Eddington limit will tend to heat and expel the gas from the vicinity
of the BH thus reducing further accretion until the gas can cool and
the cycle can start again.  There are various different time-scales
associated with these relaxation oscillations depending on whether the
heating is effective within or outside of the accretion radius.  A
small duty cycle is the natural outcome of this process in accord with
the observational fact that most elliptical galaxies are not seen in
the AGN state.  A corollary of this (low average accretion rate) is
that the black hole mass growth is self limited to relatively low
values consistent with the observed relations between BH mass and
stellar mass in ellipticals.  Another corollary is that gas, being
steadily pushed away from the dense inner part of the galaxy, radiates
far less in the X-ray bands than it would otherwise do were there no
feedback - again in accord with observations.  Finally, the QSO
luminosities are also computed to be within the observed range.

What are the critical tests of this picture that can be proposed?
First, the feedback should produce an anti-correlation between the
galaxy X-ray luminosity and the quasar luminosity: elliptical
galaxies containing QSOs should, on average, be less luminous in the X-rays
from their hot gas than the systems with the same velocity dispersion
but observed in an AGN quiet phase.
Available data may
allow one to test this prediction. 

Second, we should occasionally find ellipticals in the phase after a
series of bursts, when they contain a low density high temperature
bubble in their central regions.  Do we observe this?  Many further
such tests can be proposed by the discerning reader, and the ones
mentioned above may not, in fact, be usefully employable, but it is
certain that we have now moved from wondering about the existence of
massive black holes in the centers of galaxies to computing the
consequences of these behemoths and searching for the observational
results of the enormous energy output from them.

\section{Acknowledgements} 

In addition to the important input from our colleagues, M.G. Park,
S. Pellegrini, S. Sazonov and R. Sunyaev, we are appreciative of the
wise counsel received from J. Binney, A. Fabian, and M. Rees.

\end{document}